\documentclass{article}
\usepackage{spconf}

\usepackage[utf8]{inputenc}
\usepackage[T1]{fontenc}

\usepackage{amsmath}
\usepackage{amsfonts}

\usepackage{graphicx}
\usepackage[export]{adjustbox}

\usepackage{url}
\usepackage{times}
\usepackage{booktabs} 
\usepackage{multirow}

\usepackage{lipsum}
\usepackage{tikz}
\usetikzlibrary{calc}
\usetikzlibrary{arrows.meta}

\usepackage{cuted}
\usepackage{stfloats}
\usepackage[normalem]{ulem} 
\usepackage{bm}
\usepackage[acronym,sort=def]{glossaries}

\usepackage[switch]{lineno} 

\usepackage{todonotes}

\newcommand{\norm}[1]{\left\lVert#1\right\rVert}
\newcommand{\Enorm}[1]{\norm{#1}_2}

\newcommand{\Fnorm}[1]{\norm{#1}_F}

\newcommand{\mat}[1]{ \bm{\mathrm{#1}} }

\newcommand{\Tpose}{^{\mathrm{T}}}

\newcommand{\est}[1]{ \hat{#1} }

\newcommand{\measure}[1]{ \check{#1} }

\newcommand{\RMat}[1]{ \gls{RMat}\left(#1\right) }

\newglossaryentry{RealSet}%
{%
    name={\(\mathbb{R}\)},
    text={\mathbb{R}},
    description={Set of real number},
}
\newglossaryentry{mpVec}%
{%
    name={\( \mat{p}_{i,j} \)},
    text={\mat{p}},
    description={Pointing vector form \(i\)-th node to \(j\)-th acousitc event},
}

\newglossaryentry{mdVec}%
{%
    name={\( \mat{d}_{i,j} \)},
    text={\mat{d}},
    description={Direction of Arrival vector form \(i\)-th node to \(j\)-th acousitc event},
}
\newglossaryentry{mToFMat}%
{%
    name={\(\mat{\Psi}\)},
    text={\mat{\Psi}},
    description={Time of Flight matrix containing ToF between all nodes and all acousitc events},
}
\newglossaryentry{RMat}%
{%
    name={\(\mat{R}(\mat{\theta})\)},
    text={\mat{R}},
    description={Rotation matrix \( \mat{R} \in \mathrm{SO}(n) \) generated by \(\mat{\theta} \in \gls{RealSet}^{n-1}\)},
}
\newglossaryentry{mToF}%
{%
    name={\(\psi_{i,j}\)},
    text={\psi},
    description={Time of Flight between \(i\)-th node and \(j\)-th acousitc event},
}
\newglossaryentry{BTuple}%
{%
    name={\(\mathcal{B}\)},
    text={\mathcal{B}},
    description={Tuple of node frames},
}
\newglossaryentry{nb}%
{%
    name={\(\mat{b}_{i,k}\)},
    text={\mat{b}},
    description={\(k\)-th vector of local basis of \(i\)-th node},
}

\newglossaryentry{nB}%
{%
    name={\(\mat{B}_i\)},
    text={\mat{B}},
    description={Transformation matrix form standard basis to  basis of \(i\)-th node},
}

\newglossaryentry{eTuple}%
{%
    name={\(\mathcal{S}\)},
    text={\mathcal{S}},
    description={Tuple of event positions},
}
\newglossaryentry{eP}%
{%
    name={\(\mat{s}_j\)},
    text={\mat{s}},
    description={Vector containing position of \(j\)-th acoustic event in ambiet space},
}

\newglossaryentry{nTuple}%
{%
    name={\(\mathcal{N}\)},
    text={\mathcal{N}},
    description={Tuple of node positions},
}
\newglossaryentry{nP}%
{%
    name={\(\mat{n}_i\)},
    text={\mat{n}},
    description={Vector containing position coordiantes of \(i\)-th node in ambient space},
}

\newglossaryentry{dimas}%
{%
    name={\(D\)},
    text={D},
    description={Dimension of the ambient space},
}

\newglossaryentry{wspeed}%
{%
    name={\(c\)},
    text={c},
    description={Speed of sound},
}

\newglossaryentry{SNum}%
{%
    name={\(S\)},
    text={S},
    description={Number of acoustic events},
}

\newglossaryentry{NNum}%
{%
    name={\(N\)},
    text={N},
    description={Number of nodes},
}

\newacronym{ToF}{ToF}{Time of Flight}
\newacronym{ToA}{ToA}{Time of Arrival}
\newacronym{TDoA}{TDoA}{Time Difference of Arrival}
\newacronym{DoA}{DoA}{Direction of Arrival}

\newacronym{vMF}{vMF}{von Mises-Fisher}
\newacronym{PDF}{PDF}{Probability Density Function}
\newacronym{LS}{LS}{Least-Squares}
\newacronym{ML}{ML}{Maximum-Likelihood}

\title{%
    Exploiting rays in blind localization of distributed sensor arrays
}
\name{%
    Szymon Woźniak and Konrad Kowalczyk
    \thanks{%
        This research received financial support from the Foundation for Polish
        Science under grant number First TEAM/2017--3/23 which is co-financed by
        the European Union under the European Regional Development Fund.
    }
}
\address{%
    AGH University of Science and Technology \\
    Department of Electronics\\
    Kraków, Poland \\
    szymon.wozniak@agh.edu.pl, konrad.kowalczyk@agh.edu.pl
}

\begin{document}
\ninept{}
\maketitle
\begin{abstract}
Many signal processing algorithms for distributed sensors are capable of improving their performance if the positions of sensors are known. In this paper, we focus on estimators for inferring the relative geometry of distributed arrays and sources, i.e. the setup geometry up to a scaling factor. Firstly, we present the Maximum Likelihood estimator derived under the assumption that the Direction of Arrival measurements follow the von Mises-Fisher distribution. Secondly, using unified notation, we show the relations between the cost functions of a number of state-of-the-art relative geometry estimators. Thirdly, we derive a novel estimator that exploits the concept of rays between the arrays and source event positions. Finally, we show the evaluation results for the presented estimators in various conditions, which indicate that major improvements in the probability of convergence to the optimum solution over the existing approaches can be achieved by using the proposed ray-based estimator.
\end{abstract}
\begin{keywords}
    array processing, distributed sensor networks, geometry calibration, maximum likelihood, least squares
\end{keywords}
\section{Introduction}\label{sec:intro}

The locations of distributed devices equipped
with acoustic sensors are often ad-hoc and unknown.
In order to exploit the positioning of the distributed acoustic sensor arrays for
source localization~\cite{cobosModifiedSRPPHATFunctional2011} and speech
enhancement~\cite{kowalczykParametricSpatialSound2015,
kowalczykSoundAcquisitionNoisy2013}, their positions and
orientations need to be determined.
Different approaches like Acoustic Simultaneous Localization and Mapping \cite{ASLAM} or self-calibration methods allows to determine those unknowns.
Many self-localization methods exploit the low-rank properties of
\gls{ToF} measurements~\cite{croccoBilinearApproachPosition2012,
dokmanicEuclideanDistanceMatrices2015}.
However, in practice this requires full control over the sound sources and synchronization of capturing devices.
In~\cite{heusdensTimedelayEstimationTOAbased2014} an additional estimation of capture time offsets have been proposed using the low-rank matrix formulated in terms of \gls{ToF} measurements. In~\cite{wangSelfLocalizationAdHocArrays2016}
this method has been adapted to the \gls{TDoA} measurements. For passive distributed sensor array self-calibration, the \gls{DoA} measurements can additionally be exploited, which is typically approached in a two-step
manner.
First, the relative geometry is estimated using \gls{DoA}s, and next the absolute positions of
sensor arrays are computed based on the relative
geometry and measured \gls{TDoA}s~\cite{schmalenstroeerUnsupervisedGeometryCalibration2011,
jacobMicrophoneArrayPosition2012,jacobDOAbasedMicrophoneArray2013,
wozniakPassiveJointLocalization2019}.
A one-step approach for joint \gls{DoA}-\gls{TDoA} array self-localization has been proposed
in~\cite{plingeGeometryCalibrationMultiple2014}.

In this paper, we propose a method to find the relative geometry of several acoustic sensor arrays
and acoustic events by exploiting the concept of half-lines, also known as rays.
We formulate the \gls{DoA} based optimization problem which minimizes the error between the estimate of an acoustic event
position observed jointly by all arrays and an event position estimated by each array
independently. We show that the proposed ray-based \gls{LS} cost function is highly robust
towards random parameter initialization, which allows to avoid the usage of time-consuming methods for
finding suitable initialization parameters such as the exhaustive grid search applied e.g. in~\cite{plingeGeometryCalibrationMultiple2014}, or the multiple
initialization scheme, used e.g. in~\cite{wangSelfLocalizationAdHocArrays2016}.
In addition, we derive the \gls{ML}
estimator under the assumption that \gls{DoA} measurements follow a \gls{vMF}
distribution~\cite{mardiaDirectionalStatistics2009}. Using the unified notation, we show the relations between the presented ML estimator and a number of state-of-the-art estimators from the literature~\cite{schmalenstroeerUnsupervisedGeometryCalibration2011,
jacobMicrophoneArrayPosition2012,jacobDOAbasedMicrophoneArray2013, wozniakPassiveJointLocalization2019}.
Finally, in a set of experiments, we show the benefits of using the proposed ray-based estimator in terms of high robustness against random initialization and low variance, and compare its performance with the derived \gls{ML}
estimator and several existing estimators~\cite{schmalenstroeerUnsupervisedGeometryCalibration2011,
jacobMicrophoneArrayPosition2012,jacobDOAbasedMicrophoneArray2013,
wozniakPassiveJointLocalization2019}.

\section{Problem formulation}\label{sec:problem}

Consider a scenario in which \(\gls{NNum}\) sensor arrays, hereafter referred to as nodes, and \(\gls{SNum}\) consecutive events, emitted by one or more sources, are distributed in a \(\gls{dimas}\)-dimensional ambient space spanned by standard basis \( (\mat{e}_1, \mat{e}_2, \ldots , \mat{e}_{\gls{dimas}}) \).
The position of the \(i\)-th node is denoted as \(\gls{nP}_i \in \gls{RealSet}^{\gls{dimas}}\), while the positions of all nodes are denoted as an ordered set \(\gls{nTuple} = (\gls{nP}_1, \gls{nP}_2, \ldots, \gls{nP}_{\gls{NNum}}) \).
Similarly, the position of the \(j\)-th event is given by vector \(\gls{eP}_j \in \gls{RealSet}^{\gls{dimas}}\), and the positions of all events are grouped into an ordered set \(\gls{eTuple} = (\gls{eP}_1, \gls{eP}_2, \ldots, \gls{eP}_{\gls{SNum}}) \).
Furthermore, we assume that each node is equipped with more than one sensor and that the local geometry of sensors within each node is known \emph{a priori}, such that the \gls{DoA} can be measured at the node.
The orientation of the \(i\)-th node is defined by its local basis \((\gls{nb}_{i,1}, \gls{nb}_{i,2}, \ldots, \gls{nb}_{i,\gls{dimas}})\) which contains orthonormal vectors \( \gls{nb}_{i,d} \in \gls{RealSet}^{\gls{dimas}} \).
A change of basis from the standard basis into the local basis of the \(i\)-th node is given by a linear map
\(
    \gls{nB}_i
    =
    [
        \gls{nb}_{i,1},
        \gls{nb}_{i,2},
        \ldots,
        \gls{nb}_{i,\gls{dimas}}
    ]\Tpose \in \gls{RealSet}^{\gls{dimas} \times \gls{dimas}}
\).
In this work, we assume that the determinant of each matrix \(\gls{nB}_i\) is equal to one, and therefore those matrices are elements from the special orthogonal group, i.e., \( \gls{nB}_i \in \mathrm{SO}(\gls{dimas})\).
The ordered set containing all linear maps is denoted hereafter as \(\gls{BTuple} = (\gls{nB}_1, \gls{nB}_2, \ldots, \gls{nB}_{\gls{NNum}}) \).
The relative position of the \(j\)-th acoustic event in reference to the position of the \(i\)-th node and its local basis can be expressed by vector \(\gls{mpVec}_{i,j} \in \gls{RealSet}^{\gls{dimas}}\) defined as
\begin{equation}\label{eq:pVector}
    \gls{mpVec}_{i,j}
    =
    \gls{nB}_i \left( \gls{eP}_j - \gls{nP}_i \right).
\end{equation}
The \gls{DoA} vector \(\gls{mdVec}_{i,j} \in \gls{RealSet}^{\gls{dimas}}\) pointing towards the \(j\)-th event from the \(i\)-th node position is given by
\begin{equation}\label{eq:dVector}
    \gls{mdVec}_{i,j} = \gls{mpVec}_{i,j} \Enorm{\gls{mpVec}_{i,j}}^{-1} \; .
\end{equation}
In this work, the focus on the estimation of the so-called relative geometry, i.e. the geometry up to an unknown scaling factor, which is fully defined by three ordered sets, namely \(\gls{BTuple}\), \(\gls{nTuple}\), and
\(\gls{eTuple}\). Such a relative geometry can be estimated based solely on the \gls{DoA} measurements using the methods presented in Sec. \ref{sec:method}.
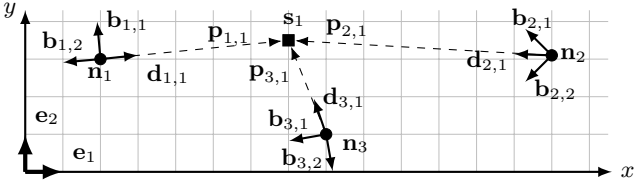
\begin{figure}\label{fig:geometry_example}
\centering
    \begin{tikzpicture}[scale=.5]
        \draw[help lines, color=gray!50] (0,0) grid (15.5,4.3);
        \draw[-latex, thick] (0,0)--(15.6,0) node[right]{\(x\)};
        \draw[-latex, thick] (0,0)--(0,4.3) node[left]{\(y\)};

        \draw[-latex, ultra thick] (0,0)--(1,0) node[above right]{$\mat{e}_1$};
        \draw[-latex, ultra thick] (0,0)--(0,1) node[above right]{$\mat{e}_2$};

        \coordinate (N1) at (2,3);
        \coordinate (N2) at (14,3.1);
        \coordinate (N3) at (8,1);
        \coordinate (S1) at (7,3.5);

        \draw[fill=black] (N1) circle (.15) node[below]{$\mat{n}_1$};
        \draw[-latex, thick,shift={(N1)},rotate=95] (0,0)--(1,0) node[right]{$\mat{b}_{1,1}$};
        \draw[-latex, thick,shift={(N1)},rotate=95] (0,0)--(0,1) node[above]{$\mat{b}_{1,2}$};
        \draw[-latex, thick,shift={(N1)},rotate=6] (0,0)--(1,0) node[below right]{$\mat{d}_{1,1}$};

        \draw[fill=black] (N2) circle (.15) node[right]{$\mat{n}_2$};
        \draw[-latex, thick,shift={(N2)},rotate=135] (0,0)--(1,0) node[above,xshift=1mm,yshift=-1.5mm]{$\mat{b}_{2,1}$};
        \draw[-latex, thick,shift={(N2)},rotate=135] (0,0)--(0,1) node[below right,xshift=0mm,yshift=1.5mm]{$\mat{b}_{2,2}$};
        \draw[-latex, thick,shift={(N2)},rotate=178] (0,0)--(1,0) node[left,xshift=0.5mm,yshift=-1mm]{$\mat{d}_{2,1}$};

        \draw[fill=black] (N3) circle (.15) node[right, ,xshift=1mm,yshift=-1.5mm]{$\mat{n}_3$};
        \draw[-latex, thick,shift={(N3)},rotate=-170] (0,0)--(1,0) node[above]{$\mat{b}_{3,1}$};
        \draw[-latex, thick,shift={(N3)},rotate=-170] (0,0)--(0,1) node[left,xshift=0mm,yshift=1.5mm]{$\mat{b}_{3,2}$};
        \draw[-latex, thick,shift={(N3)},rotate=109] (0,0)--(1,0) node[right]{$\mat{d}_{3,1}$};

        \draw[fill=black,shift={($(S1)-(0.15,0.15)$)}] (0,0) rectangle (0.3, 0.3) node[above]{$\mat{s}_1$};

        \draw[-Latex,dashed, shorten >= 0.9mm] (N1) -- (S1) node[pos=0.8, left,xshift=1mm,yshift=1mm]{$\mat{p}_{1,1}$};
        \draw[-Latex,dashed, shorten >= 0.9mm] (N2) -- (S1) node[pos=0.8, above,xshift=0.8mm,yshift=-0.5mm]{$\mat{p}_{2,1}$};
        \draw[-Latex,dashed, shorten >= 0.9mm] (N3) -- (S1) node[pos=0.7,left,yshift=-0.9mm]{$\mat{p}_{3,1}$};
    \end{tikzpicture}
    \caption{%
        Example geometry with three nodes and one event in a
        two-dimensional ambient space.
        Dashed vectors denote \(\gls{mpVec}_{i,j}\).
    }
\end{figure}

\section{Blind estimation of the relative geometry}%
\label{sec:method}

This section describes several estimators of \(\gls{BTuple}\), \(\gls{nTuple}\), and \(\gls{eTuple}\), which fully define the relative geometry.
In Sec.~\ref{subsec:method_ML} we derive an \gls{ML} estimator using \gls{vMF} \gls{PDF}. Then, using cosine angular distance, we show the relations between the presented \gls{ML} estimator and several estimators known from the literature. In Sec.~\ref{subsec:method_rays}, a novel \gls{LS} estimator is proposed, in which the ray equation is exploited instead of the angular distance. Finally, Sec.~\ref{subsec:scaling} discusses the most robust and accurate approaches and presents our suggested processing chain.

\subsection{Maximum likelihood estimator and relation to prior work}%
\label{subsec:method_ML}

Let us assume that the measured \gls{DoA}s given by unit-length vectors
\(\measure{\gls{mdVec}}_{i,j}\) follow
\gls{vMF} distribution in a D-dimensional space~\cite{mardiaDirectionalStatistics2009}, with operator \(\measure{(\cdot)}\) denoting the measured value.
In this case, the \gls{PDF} of the measured \gls{DoA}s is given by
\begin{equation}\label{eq:pdf_vMF}
    f \left(
        {\gls{mdVec}}_{i,j}; \; \measure{\gls{mdVec}}_{i,j}, \kappa_{i,j}
    \right)
    =
    Z(\kappa_{i,j}) \;
    e^{\kappa_{i,j} \, \gls{mdVec}_{i,j}\Tpose \, \measure{\gls{mdVec}}_{i,j}}
    ,
\end{equation}
where \(Z(\kappa_{i,j})\) is the normalization factor given by
\begin{equation}
    Z(\kappa_{i,j})
    =
        (\kappa_{i,j})^{\frac{\gls{dimas}}{2}-1}
        \left(
            (2\pi)^{\frac{\gls{dimas}}{2}} \,
            I_{\frac{\gls{dimas}}{2}-1} (\kappa_{i,j})
        \right)^{-1}
    \, ,
\end{equation}
\(\kappa_{i,j}\) denotes the concentration parameter of the \gls{PDF}
around the mean direction \(\gls{mdVec}_{i,j} \), while \(I_m(\kappa)\) is
the \(m\)-th order modified Bessel function of the first kind.
Assuming homoscedasticity of DoA measurements,
we set \(\kappa_{i,j}=1\) for all \((i,j)\) pairs.
Under this assumption, the \gls{ML} estimation problem can be formulated as
\begin{equation}
\label{eq:vMF_constrained}
    \est{\gls{BTuple}}, \tilde{\gls{nTuple}}, \tilde{\gls{eTuple}}
    =
    \underset{\gls{BTuple}, \gls{nTuple}, \gls{eTuple}}{\mathrm{argmax}}
    \sum_{i=1}^{\gls{NNum}} \sum_{j=1}^{\gls{SNum}}
    \gls{mdVec}_{i,j}\Tpose \; \measure{\gls{mdVec}}_{i,j}
    \;\;\;\text{s.t.}\;\;\; \gls{nB}_i \in \mathrm{SO}(\gls{dimas})\;.
\end{equation}
The constraint on the linear maps can be removed by using the rotation
matrices \(\mat{B}_i = \mat{R}(\mat{\theta}_i)\) with generators given by vector \(\mat{\theta}_i\).
Consequently, instead of searching for the entire set \(\gls{BTuple}\), it is sufficient to find unknown generators
\(\mat{\Theta} = [\mat{\theta}_1, \dots, \mat{\theta}_{\gls{NNum}}]\) of
\(\gls{BTuple}\).
The final \gls{ML} estimation of the relative geometry based on the \gls{vMF} distribution for DoAs under the homoscedasticity assumption can be expressed by
\begin{equation}\label{eq:loss_vMF}
    \est{\mat{\Theta}}, \tilde{\gls{nTuple}}, \tilde{\gls{eTuple}}
    =
    \underset{\mat{\Theta}, \gls{nTuple}, \gls{eTuple}}{\mathrm{argmax}}
    \sum_{i=1}^{\gls{NNum}} \sum_{j=1}^{\gls{SNum}}
    J_{i,j} \; ,
\end{equation}
where \(J_{i,j}\) is the cosine of the angle between the measured and estimated
\gls{DoA} vectors, which is given by
\begin{equation}\label{eq:angular_dist}
    J_{i,j}
    =
    \gls{mdVec}_{i,j}\Tpose \; \measure{\gls{mdVec}}_{i,j}
    =
    \frac{
        \left(\gls{nP}_i - \gls{eP}_j\right)\Tpose
    }{
        \Enorm{\gls{eP}_j - \gls{nP}_i}
    }
    \left\{\RMat{\mat{\theta}_i}
    \right\}\Tpose \measure{\gls{mdVec}}_{i,j} \; .
\end{equation}

Over the years, several different DoA-based estimators for the relative geometry estimation have been proposed in the literature, among others \cite{schmalenstroeerUnsupervisedGeometryCalibration2011, jacobMicrophoneArrayPosition2012,  jacobDOAbasedMicrophoneArray2013, wozniakPassiveJointLocalization2019}. In the following, using the proposed unified notation, we present the mathematical relations between these approaches in terms a cosine angular distance \(J_{i,j}\) as an attempt to present to the reader the similarities and differences between the existing approaches.
In \cite{schmalenstroeerUnsupervisedGeometryCalibration2011}, the optimization problem has been formulated as a system of non-linear equations and later rewritten to simpler form by \cite{jacobMicrophoneArrayPosition2012}. The resulting estimator \cite{schmalenstroeerUnsupervisedGeometryCalibration2011} can be rewritten as
\begin{equation}\label{eq:loss_Schmalen2011}
    \est{\mat{\Theta}}, \tilde{\gls{nTuple}}, \tilde{\gls{eTuple}}
    =
    \underset{\mat{\Theta}, \gls{nTuple}, \gls{eTuple}}{\mathrm{argmin}}
    \sum_{i=1}^{\gls{NNum}} \sum_{j=1}^{\gls{SNum}}
    \gls{mToF}_{i,j}^2
    \left(
        1 - J^2_{i,j}
    \right) \; ,
\end{equation}
In \cite{jacobMicrophoneArrayPosition2012} the estimator has been derived based on geometric relations, and its cost function can be expressed using a cosine distance as
\begin{equation}\label{eq:loss_Jacob12}
    \est{\mat{\Theta}}, \tilde{\gls{nTuple}}, \tilde{\gls{eTuple}}
    =
    \underset{\mat{\Theta}, \gls{nTuple}, \gls{eTuple}}{\mathrm{argmin}}
    \sum_{i=1}^{\gls{NNum}} \sum_{j=1}^{\gls{SNum}}
    \gls{mToF}_{i,j}^2
    \left( 1-J_{i,j} \right)^2
\end{equation}
where \(\gls{mToF}_{i,j}\) denotes a heuristically set weight that is equal to the distance between the \(j\)-th event and the \(i\)-th node, i.e., \(\gls{mToF}_{i,j} = \Enorm{\gls{mpVec}_{i,j}}\).
On the other hand, the estimator proposed in~\cite{jacobDOAbasedMicrophoneArray2013} has been derived using circular statistics and some heurestic weight was also advised as in \cite{jacobMicrophoneArrayPosition2012}, which results in the following cost function
\begin{equation}\label{eq:loss_Jacob13}
    \est{\mat{\Theta}}, \tilde{\gls{nTuple}}, \tilde{\gls{eTuple}}
    =
    \underset{\mat{\Theta}, \gls{nTuple}, \gls{eTuple}}{\mathrm{argmin}}
    \sum_{i=1}^{\gls{NNum}} \sum_{j=1}^{\gls{SNum}}
    \gls{mToF}_{i,j}
    \left( 1-J_{i,j} \right) \;,
\end{equation}
Recently, a LS estimator with the cost function defined as a sum of Euclidean distances between the measured and estimated \gls{DoA}s has been proposed in~\cite{wozniakPassiveJointLocalization2019}, and it can be written as
\begin{equation}\label{eq:loss_Wozniak2019}
    \est{\mat{\Theta}}, \tilde{\gls{nTuple}}, \tilde{\gls{eTuple}}
    =
    \underset{\mat{\Theta}, \gls{nTuple}, \gls{eTuple}}{\mathrm{argmin}}
    \sum_{i=1}^{\gls{NNum}} \sum_{j=1}^{\gls{SNum}}
    \frac{
        \left( 1-J_{i,j} \right)
    }{2} \; .
\end{equation}
Note that the minimization of the cost function given by \eqref{eq:loss_Wozniak2019} is equivalent to the maximization of the \gls{ML} cost function given by \eqref{eq:loss_vMF}.
On the other hand, the only difference between the ML cost function given by \eqref{eq:loss_Wozniak2019} and cost function \eqref{eq:loss_Jacob13} is a heuristically set multiplicative weight \(\gls{mToF}_{i,j}\).
In addition, it can be noticed that the difference between cost function \eqref{eq:loss_Jacob12} and cost function \eqref{eq:loss_Jacob13} is only in squaring the elements of the sum.
Note that cost function \eqref{eq:loss_Schmalen2011} as presented in \cite{jacobMicrophoneArrayPosition2012} can be expressed in terms of squared sinuses, and its relation to the ML cost function can be shown by using the Pythagorean trigonometric identity.

\subsection{The proposed estimator based on the ray equation}%
\label{subsec:method_rays}

In this section, we derive a novel estimator, which is based on the ray (i.e., half-line) equation. We begin by rewriting~\eqref{eq:pVector} to obtain an expression for the position of the \(j\)-th event,
and factor out the scalar value \(\Enorm{\gls{mpVec}_{i,j}}\), which yields the following expression:
\begin{equation}\label{eq:preray}
    \gls{eP}_j
    =
    \gls{nB}_i\Tpose
    \Enorm{\gls{mpVec}_{i,j}}
    \frac{\gls{mpVec}_{i,j}}{{}_{\phantom{2}}\Enorm{\gls{mpVec}_{i,j}}}
    +
    \gls{nP}_i\;.
\end{equation}
The first additive term in \eqref{eq:preray} can be decomposed into two components, namely the distance between the \(j\)-th event and \(i\)-th node positions, denoted as \(\gls{mToF}_{i,j} = \Enorm{\gls{mpVec}_{i,j}}\), and the direction vector \(\gls{mdVec}_{i,j}\) defined as in~\eqref{eq:dVector}. Next, we can enforce the distance parameter \(\gls{mToF}_{i,j}\) to be non-negative and write a general half-line equation
\begin{equation}\label{eq:ray}
    \gls{eP}_j^{(i)}
    =
    \gls{mToF}_{i,j}
    \gls{nB}_i\Tpose
    {\gls{mdVec}}_{i,j}
    +
    \gls{nP}_i,
    \quad \quad \quad
    \psi_{i,j} \ge 0 \, ,
\end{equation}
which describes a ray pointing
from the position of the \(i\)-th node \(\gls{nP}_i\) in the direction defined by the rotated vector
\(\gls{nB}_i\Tpose\gls{mdVec}_{i,j}\), and which intersects with the point \(\gls{eP}_j\) at a specific value of parameter \(\gls{mToF}_{i,j}\).

Next, we can exploit the fact that for the \(j\)-th event, the rays generated based on the \gls{DoA}s measured in all nodes should intersect in a common point. However, due to the measurement noise and room reverberation, a single point of intersection is often not achieved in practice.
In order to obtain an estimate of the \(j\)-th event position using
the information provided by all nodes, we propose to use the \gls{ML} estimator
under the assumption that each position \(\gls{eP}_j^{(i)}\) is corrupted by an
isotropic Gaussian noise. For a single event, the resulting estimator is given by
\begin{equation}\label{eq:event_pos_aver}
    \bar{\gls{eP}}_j
    =
    \frac{1}{\gls{NNum}}
    \sum^{\gls{NNum}}_{i=1}
    \left(
        \gls{mToF}_{i,j}
        \gls{nB}_i\Tpose
        {\gls{mdVec}}_{i,j}
        +
        \gls{nP}_i
    \right).
\end{equation}
\begin{figure*}[!b]
\begin{equation}\label{eq:loss_ray2}\tag{16}
    \est{\bm{\Theta}}, \tilde{\gls{nTuple}}, \tilde{\gls{mToFMat}}
    =
    \underset{\bm{\Theta}, \gls{nTuple}, \gls{mToFMat}}{\mathrm{argmin}}
    \sum^{\gls{SNum}}_{j=1} \sum^{\gls{NNum}}_{i=1}
    \Enorm{%
        \left(
            \gls{mToF}_{i,j}
            \left\{\mat{R}(\mat{\theta}_i)\right\}\Tpose
            \measure{\gls{mdVec}}_{i,j}
            -
            \frac{1}{\gls{NNum}} \sum^{\gls{NNum}}_{k=1}
            \gls{mToF}_{k,j}
            \left\{\mat{R}(\mat{\theta}_k)\right\}\Tpose
            \measure{\gls{mdVec}}_{k,j}
        \right)
        +
        \left(
            \gls{nP}_i - \frac{1}{\gls{NNum}} \sum^{\gls{NNum}}_{k=1} \gls{nP}_k
        \right)
    }^2
    \;\;\; \text{s.t.}\;\;\;
    \psi_{i,j} \ge \lambda
\end{equation}
\end{figure*}
Based on~\eqref{eq:ray} and~\eqref{eq:event_pos_aver}, we can formulate
the constrained \gls{LS} problem for joint estimation
of the entire relative geometry, including all node and event positions. The residual is defined as the distance between the position of the \(j\)-th
event observed at each array \(\gls{eP}_j^{(i)}\) and the \gls{ML} estimate
of the event position \(\bar{\gls{eP}}_j\). Minimization of the aforementioned residuals performed subject to two constraints, namely (i) the distance value \(\gls{mToF}_{i,j}\) has to be real and non-negative,
and (ii) the linear map \(\gls{nB}_i\) has to belong
to the special orthonormal group for the \(D\)-dimensional Euclidean space. Thus the proposed optimization problem can be written as
\begin{multline}\label{eq:loss_ray}
    \est{\gls{BTuple}}, \tilde{\gls{nTuple}}, \tilde{\gls{mToFMat}}
    =
    \underset{\gls{BTuple}, \gls{nTuple}, \gls{mToFMat}}{\mathrm{argmin}}
    \sum^{\gls{SNum}}_{j=1} \sum^{\gls{NNum}}_{i=1}
    \Enorm{%
        \gls{eP}_j^{(i)} - \bar{\gls{eP}}_j
    }^2
    \\
    \;\text{subject to}\;\;\;
    \gls{mToF}_{i,j} \ge 0
    \;\;\;\text{and}\;\;\;
    \gls{nB}_i \in \mathrm{SO}(\gls{dimas}).
\end{multline}
The cost function~\eqref{eq:loss_ray} is prone to inherently converging
to a solution from the null space.
The reason for that is that the value of the cost function is strongly
dependent on unknown distances \(\gls{mToF}_{i,j}\).
When minimizing the cost function, the distances get minimized and
the positions of nodes converge to a common location.
To avoid solutions from the null space, we constrain the distances to be greater than zero by introducing a small real positive number \(\lambda\). Similarly to \eqref{eq:vMF_constrained}, instead of estimating the set \(\gls{BTuple}\), we estimate the set \(\mat{\Theta}\) that generates \(\gls{BTuple}\) using rotation matrices.
The final cost function for the LS estimation of the relative geometry based on the ray equations is formulated by substituting~\eqref{eq:ray}
and~\eqref{eq:event_pos_aver} into~\eqref{eq:loss_ray}, which yields Equation~\eqref{eq:loss_ray2} provided at the bottom of the page.
\setcounter{equation}{16}

\subsection{The proposed refined method}%
\label{subsec:scaling}

All in all, the relative geometry can be found using any of the estimators presented in Secs. \ref{subsec:method_ML} and \ref{subsec:method_rays}.
However, each of them has different properties in terms of variance and robustness against random initialization in convergence to the global minimum.
As confirmed by the experimental evaluation shown in Sec.~\ref{sec:evaluation}, we propose to use the novel ray-based estimator given by~\eqref{eq:loss_ray2} due to its low probability of getting trapped in local minima and almost the lowest variance.
To increase the accuracy of the estimated relative geometry, we propose to refine
the results obtained using~\eqref{eq:loss_ray2} by running the ML estimator given
by~\eqref{eq:loss_Wozniak2019}
or it's modified version proposed in~\cite{wozniakSelfLocalizationDistributedMicrophone2019}.

\section{Absolute geometry estimation}\label{sec:absolute}

Having found the relative geometry of the node and source event positions, the absolute geometry is computed by scaling the relative geometry by an unknown scale parameter \(\gamma\) in the following manner
\(\est{\gls{nP}}_{i} = \est{\gamma} \; \tilde{\gls{nP}}_i\) and \(\est{\gls{eP}}_j = \est{\gamma} \; \tilde{\gls{eP}}_j\).
Typically such estimation is based on the \gls{TDoA} measurements.
To this end, a simple averaging estimator based on inter-array \gls{TDoA}s has been presented in \cite{schmalenstroeerUnsupervisedGeometryCalibration2011,jacobMicrophoneArrayPosition2012}.
Recently, an efficient \gls{LS} estimator of \(\gamma\) has been derived in \cite{wozniakPassiveJointLocalization2019}, which additionally finds onset recording times for inter-array synchronization.
Another approach presented in \cite{jacobDOAbasedMicrophoneArray2013} exploits intra-array \gls{TDoA}s but its effectiveness is limited to the cases when events are located in close proximity of the arrays.
In this work, for evaluation we use the method presented in \cite{wozniakPassiveJointLocalization2019}.

\section{Results and Evaluation}\label{sec:evaluation}

Throughout the experiments, we evaluate the cost functions given by \eqref{eq:loss_Schmalen2011},~\eqref{eq:loss_Jacob12},~\eqref{eq:loss_Jacob13},~\eqref{eq:loss_Wozniak2019}, and~\eqref{eq:loss_ray2} in a room of size \(10 \times 10 \times 3\)~[m].
To minimize those cost function, we use an optimization algorithm described in~\cite{byrdLimitedMemoryAlgorithm1995}.
The first experiment evaluates the success ratio of the presented cost functions in reaching a global minimum with random parameter initialization and perfect DoA and TDoA measurements.
To this end,
we adopt the strategy from \cite{dokmanicEuclideanDistanceMatrices2015} and consider a success when the following conditions are jointly met:
\(
    \varepsilon(\gls{BTuple}, \est{\gls{BTuple}}) \le 0.01 \;\text{and}\;
    \varepsilon(\gls{nTuple}, \est{\gls{nTuple}}) \le 0.01 \;\text{and}\;
    \varepsilon(\gls{eTuple}, \est{\gls{eTuple}}) \le 0.01
\)
, where
\begin{equation}
    \varepsilon(\mathcal{X}, \est{\mathcal{X}})
    =
    \frac{%
        \sum_{i=1}^{|\mathcal{X}|} \Fnorm{\est{x}_i - x_i}
    }{%
        \sum_{i=1}^{|\mathcal{X}|}  \Fnorm{x_i}
    }\;,
\end{equation}
denotes the error between a set of ground truth parameters \(x_i \in \mathcal{X}\) and a set of estimated parameters \(\est{x}_i \in \est{\mathcal{X}}\),
\(\Fnorm{\cdot}\) is the Frobenius norm, and \(|\mathcal{X}|\) is a number of elements in set \(\mathcal{X}\).
\begin{figure*}[!t]
    \centering
    \includegraphics[width=0.999\linewidth]{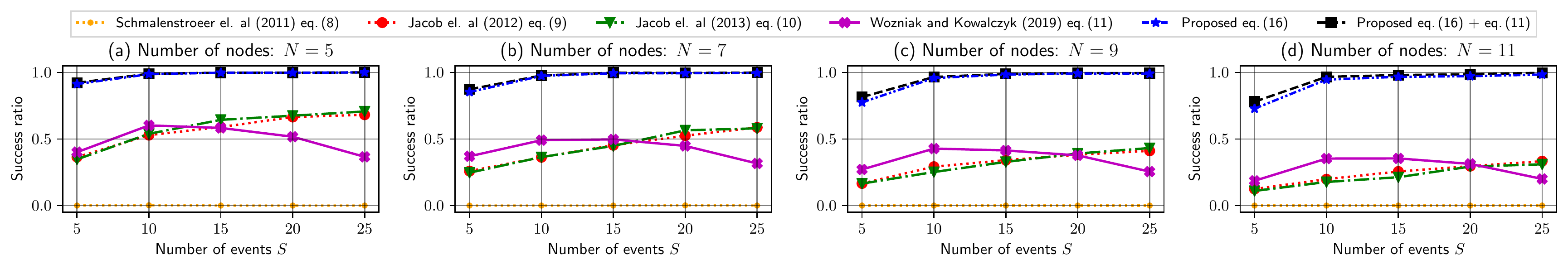}
    \vspace*{-8mm}
    \caption{%
        Success ratio of different estimators in convergence to the global minimum
        with noiseless DoAs and random initial parameters, for the varying number of nodes and events.
        For every test point we generated 1000 random realization of the network geometry.
    }
    \label{fig:exp_conv}
    \vspace{-2mm}
\end{figure*}
Figure~\ref{fig:exp_conv} presents the success ratios of different cost functions for the varying number of nodes and events with random realization of geometry for each test.
As can be observed, almost all cost functions improve their success ratios with the increasing number of events, except for \eqref{eq:loss_Wozniak2019} which tends to reach its best success ratio for the specific number of events for a given number of nodes.
The cost functions that are based on the angular distance exhibit significantly lower success ratio than the proposed ray-based cost function \eqref{eq:loss_ray2}.
Its success ratio tends to quickly decrease with an increasing number of nodes, yet the proposed ray-based cost function maintains a very high success ratio exceeding 0.95 value for \(\gls{SNum} \geq 10\).
The proposed refinement of estimates obtained with \eqref{eq:loss_ray2} using \eqref{eq:loss_Wozniak2019} leads to a minor increase of the success ratio, which indicates that the estimates obtained from the local minimum of \eqref{eq:loss_ray2} usually lead \eqref{eq:loss_Wozniak2019} to the local minimum as well.
The almost zero success ratio of \eqref{eq:loss_Schmalen2011} is caused by ambiguity of this cost function for orientation of node (i.e., condition \(\varepsilon(\gls{BTuple}, \est{\gls{BTuple}}) \le 0.01\) is not met), we referee to \cite{jacobMicrophoneArrayPosition2012} for details.
\begin{figure*}[!t]
    \centering
    \includegraphics[width=0.999\linewidth]{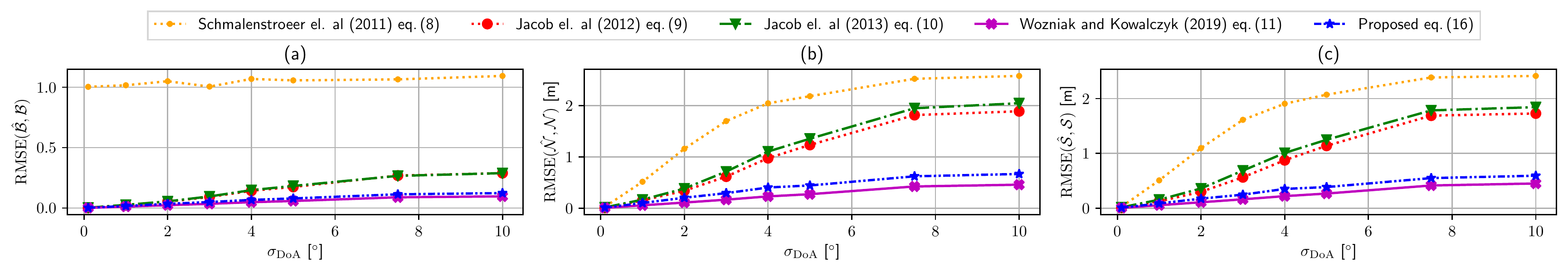}
    \vspace*{-8mm}
    \caption{%
        RMSE errors of different estimators for noisy DoAs with increasing standard deviation and ground truth
        initialization of parameters.
        For every test point we generated 1000 different realizations of the network geometry with \(\gls{NNum}=5\) and \(\gls{SNum}=10\).
    }\label{fig:exp_variance}
    \vspace{-2mm}
\end{figure*}

The second experiment evaluates the variances of estimates, and it is performed for \(\gls{NNum}=5\) and \(\gls{SNum}=10\) with an increasing standard deviation \(\sigma_{\mathrm{DoA}}\) of DoAs and ground truth values of \gls{TDoA}s.
For each cost function, the optimizer is initialized with the ground truth values of parameters in order to avoid getting trapped in local minima.
Figure~\ref{fig:exp_variance} presents the resulting Root Mean Square Errors (RMSE) of the estimated parameters.
As can be seen, the RMSE errors of all parameters are steadily increasing with an increasing \(\sigma_{\mathrm{DoA}}\).
The presented \gls{ML} cost function \eqref{eq:loss_Wozniak2019} and the proposed ray-based cost function \eqref{eq:loss_ray2} significantly outperform other methods, while \eqref{eq:loss_ray2}
is slightly less accurate than \eqref{eq:loss_Wozniak2019}.
Thus the subsequent step after the ray-based relative geometry estimation should be to perform the refining estimation using cost function \eqref{eq:loss_Wozniak2019} when more accurate relative geometry estimate is sought for.
We also report that the following cost functions \eqref{eq:loss_Schmalen2011}, \eqref{eq:loss_Jacob12}, \eqref{eq:loss_Jacob13}, are sometimes unstable during numerical optimization due to the factor \(\gls{mToF}_{i,j}\) that leads them to the solutions from the null space. We have not found any effective method to prevent this behavior without getting worse results.

The aim of the third experiment is to evaluate the proposed ray-based estimation in more realistic acoustic conditions in a reverberant room. Room impulse responses in two rooms with reverberation times of \(\text{RT}_{60} = \{400, 800\}\)~ms are simulated using the image-source method~\cite{allenImageMethodEfficiently1979} for \(\gls{NNum}=5\) and \(\gls{SNum}=10\). Each node consists of 8 microphones arranged on a cube with edge's length of 0.1~m, and SRP-PHAT ~\cite{dibiaseRobustLocalizationReverberant2001} and GCC-PHAT~\cite{gccphat} are used to estimate DoAs and TDoAs, respectively. In this experiment, we do not evaluate the following cost functions: \eqref{eq:loss_Schmalen2011}, \eqref{eq:loss_Jacob12}, and \eqref{eq:loss_Jacob13} since in practice their success ratio is low and the RMSE results would be heavily biased by the results obtained when the cost functions got trapped in local minima.
These cost functions would require the application of additional methods for optimization which are often based on multiple realization conditions such as the Random Sample Consensus (RANSAC) method \cite{fischler1981random} which has been applied e.g. in \cite{schmalenstroeerUnsupervisedGeometryCalibration2011, jacobMicrophoneArrayPosition2012,jacobDOAbasedMicrophoneArray2013} or multiple-time initialization scheme applied in \cite{wangSelfLocalizationAdHocArrays2016}.
Note that in general RANSAC could also be applied to the proposed estimation chain in order to obtain even better results.
Table~\ref{tab:room_results} presents the RMSE results for the evaluated cost functions.
The results of experiments performed in reverberant rooms indicate that the proposed refined method indeed offers an improvement in the relative geometry estimation, yet the robustness against random initialization achieved by \eqref{eq:loss_ray2} is preserved.

\begin{table}
    \footnotesize{}
    \centering
    \caption{RMSE of estimated parameters for 100 random geometries in simulated reverberant rooms of size \(10\times{}10\times{}3\) m with \(\text{RT}_{60}=400\) and \(800\)~ms.}
    \label{tab:room_results}
    \begin{tabular}{ccccc}
        Cost &
        \(\text{RT}_{60}\) &
        \(\mathrm{RMSE}(\est{\gls{BTuple}})\) &
        \(\mathrm{RMSE}(\est{\gls{nTuple}})\) &
        \(\mathrm{RMSE}(\est{\gls{eTuple}})\) \\ \toprule
        \multirow{2}{*}{\eqref{eq:loss_ray2}}
        & 400 [ms] & 0.09 & 0.21 [m] & 0.26 [m] \\ \cmidrule{2-5}
        & 800 [ms] & 0.28 & 0.53 [m] & 0.67 [m] \\ \midrule
        \multirow{2}{*}{\eqref{eq:loss_ray2}+\eqref{eq:loss_Wozniak2019}}
        & 400 [ms] & 0.03 & 0.11 [m] & 0.15 [m]  \\ \cmidrule{2-5}
        & 800 [ms] & 0.12 & 0.29 [m] & 0.39 [m]  \\ \bottomrule
    \end{tabular}
\end{table}

\section{Conclusions}\label{sec:conclusion}
We have proposed an estimator of the relative geometry of
distributed sensor arrays and events that is based on the ray equation.
Furthermore, we derive the ML estimator based on the \gls{vMF} \gls{PDF} and relate to prior work.
The results of performed experiments indicate that the proposed ray-based estimator is highly robust towards random initialization and that the ML estimator is highly accurate in position estimation in comparison with state-of-the-art approaches.

\bibliographystyle{IEEEtran}

\end{document}